\documentclass[aps,pre,showkeys,preprint
]{revtex4}
\usepackage{graphicx}
\usepackage{amsmath}
\usepackage{amsfonts}
\usepackage{amssymb}
\usepackage{bm}
\usepackage{enumitem}  
\usepackage{float}
\usepackage{xcolor}
\usepackage{color}

\begin{document}


\title{Exchange fluid model derived from first principles quantum kinetic theory for plasmas}

\author{Fernando Haas}
\affiliation{
Physics Institute, Federal University of Rio Grande do Sul, Av. Bento Gon\c{c}alves 9500, 91501-970 Porto Alegre, RS, Brazil\\
Email Address: fernando.haas@ufrgs.br}

\begin{abstract}

Starting from first principles quantum kinetic theory for ideal plasmas with exchange effects, the quantum hydrodynamic equations are derived taking moments of the corresponding exchange-Vlasov equation. The case of an electron-ion plasma where ions are entirely classical is considered. The linear dispersion relation for ion-acoustic waves is found from the macroscopic equations and compared with exchange quantum kinetic theory, yielding a qualitative agreement apart from a numerical factor of order one in the exchange contribution, assuming a Maxwellian background as a first step, for analytical simplicity. The validity conditions of the treatment are discussed and exchange effects are shown to be necessarily a correction, within the ideal and long wavelength approximations. 

\end{abstract}

\maketitle

\section{Introduction}

The exchange energy arises from the antisymmetrization of the
electron wavefunctions considered in the average electrostatic interaction energy. This
exchange energy (per unit volume) was first
evaluated by Bloch \cite{Bloch}, in terms of a $\sim n^{4/3}$ contribution, where $n$ is the electrons number density. 
Inclusion of exchange along the Thomas-Fermi model leads to the Thomas-Fermi-Dirac equation. The overall effect of the exchange
is to lower the energy and pressure of the Thomas-Fermi model \cite{Eliezer}.

Recently, exchange effects have been incorporated in quantum kinetic theory for plasmas \cite{Zamanian}, deriving the Wigner equation for electrons
within the Hartree-Fock approximation, taking into account the complete antisymmetry of the N-particle electrons density matrix. It was assumed that the plasma is not spin polarized and that the relevant length scales are much longer than the
thermal de Broglie wavelength. This later approximation means that quantum diffraction effects are not taken into account. The exchange kinetic theory was applied to ion-acoustic and Langmuir waves in completely degenerate plasmas \cite{Ekman}, showing exact agreement with previous results from another methods \cite{Roos}-\cite{Nozieres}. A generalization of exchange kinetic theory to electromagnetic (non electrostatic) phenomena has been proposed \cite{Zamanian2}. The impact of exchange on linear plasma waves has always been treated perturbatively. 

It happens that the resulting kinetic equation - the Vlasov equation modified by exchange effects - is quite complicated, see Eq. (\ref{e1}) below, which makes its applicability limited to some extent. Indeed it is a cumbersome nonlinear integro-differential equation in spite of the approximations made. In practice, nonlinear phenomena are outside the scope of the exchange kinetic theory. At the present, not even numerical simulations of the exchange kinetic equation have been made, which could be useful for nonlinear waves. Nevertheless, the first principles new kinetic equation can be used for the validation for other calculation schemes, like density functional theory, for instance. It is clearly desirable to have simplified models starting from the basic kinetic exchange equation, which are at the same time more amenable to analytical and nonlinear approaches. 

In this work we derive the macroscopic model based on the moments of the exchange kinetic equation. The moments
approach is traditional in classical kinetic theory \cite{Grad}. It has been also applied for quantum plasmas, either electrostatic \cite{PLA} or electromagnetic in a gauge invariant setting \cite{NJP}. In contrast, exchange-correlation effects have been incorporated \cite{Crouseilles} in fluid models for plasmas in terms of effective potentials derived \cite{Brey} in density functional theory (DFT), becoming a popular approach \cite{Xia}-\cite{Zeba}. However, the exchange fluid models with effective DFT potentials have been examined, with the conclusion that they have a good agreement with exchange kinetic theory for large phase velocities in comparison with the Fermi velocity in degenerate plasmas, becoming less satisfactory for shorter wavelengths. For instance, for high
frequency waves, the DFT-based hydrodynamical model predicts the wrong sign of the exchange contribution, for short wavelengths \cite{Brodin2019}. For these reasons, it is an essential task, to derive exchange macroscopic models from first principles, starting right from the kinetic theory and evaluating the moments dynamics. The price for the choice of the moments method focused on the time-evolution of average quantities like number density, fluid velocity field, pressure dyad etc. is the loss of information on kinetic aspects, related to Landau damping, the plasma echo and so on.  

As will be discussed, even in the macroscopic approach the exchange effects still have a prominent influence of the underlying kinetic distribution function. As a first exploration, here the exchange effects on ion-acoustic waves are studied from a fluid exchange model assuming a background Maxwellian equilibrium. This choice is the same as in the exchange kinetic treatment in Ref. \cite{Zamanian}, since in this case analytical results are more accessible than for a Fermi-Dirac equilibrium, for instance. In this way we can have a detailed account on the similarities and differences between the macroscopic and microscopic approaches. 

The article is organized as follows. In Sec. II,  the exchange fluid equations are derived from the moments of the exchange kinetic equation in the electrostatic limit. In Sec. III, we consider the
impact on ion-acoustic waves by treating the exchange effects
perturbatively within the linear approximation. Section IV discuss the validity conditions of the present approach. Section V compares the results from exchange kinetic and hydrodynamic moments theories, regarding the ion-acoustic wave in a Maxwellian background. Section VI contains the conclusions. 

\section{Exchange fluid equations}

The starting point is the kinetic equation derived in \cite{Zamanian}, see also \cite{Zamanian2}, which is the evolution equation for the electron Wigner function $f = f({\bf x}, {\bf p}, t)$ supposing the two-particle density matrix as an antisymmetric product of one-particle
density matrices. In the absence of spin polarization, neglecting correlation and quantum diffraction effects, and in the long wavelength approximation, the kinetic equation is 
\begin{eqnarray}
\frac{\partial}{\partial t}\!\!\!\!\!&f&\!\!\!\!\!({\bf x}, {\bf p}, t) + \!\! \frac{{\bf p}}{m}\cdot\nabla f({\bf x}, {\bf p}, t) - e{\bf E}({\bf x},t)\cdot\frac{\partial}{\partial{\bf p}} f({\bf x}, {\bf p}, t) \nonumber \\ \label{e1} &=& \frac{1}{2}\frac{\partial}{\partial p_i}\int\!\! d^{3}r\, d^{3}q\, e^{-i{\bf r}\cdot{\bf q}/\hbar}\,\frac{\partial V({\bf r})}{\partial r_i}\,f\left({\bf x}-\frac{\bf r}{2}, {\bf p}+\frac{\bf q}{2}, t\right)f\left({\bf x}-\frac{\bf r}{2}, {\bf p}-\frac{\bf q}{2}, t\right)  \\ &-& \!\!\frac{i\hbar}{8}\frac{\partial}{\partial p_i}\frac{\partial}{\partial p_j}\!\int\!\! d^{3}r\, d^{3}q\, e^{-i{\bf r}\cdot{\bf q}/\hbar}\,\frac{\partial V({\bf r})}{\partial r_i}\!\left[f\left({\bf x}\!-\!\frac{\bf r}{2},{\bf p}\!-\!\frac{\bf q}{2},t\right)\!\left(\!\frac{\overleftarrow{\partial}}{\partial x_j} - \frac{\overrightarrow{\partial}}{\partial x_j}\!\right)\!f\left({\bf x}\!-\!\frac{\bf r}{2},{\bf p}\!+\!\frac{\bf q}{2},t\right)\right] , \nonumber 
\end{eqnarray}
where $V({\bf r}) = e^{2}/(4\pi\varepsilon_0 r)$ and the symbols have their usual meaning \cite{Zamanian, Zamanian2}. Summation over repeated indexes is assumed. Equation (\ref{e1}) keeps the bare exchange effects, which presently are our focus.

To proceed to evolution equations for macroscopic quantities, it is necessary to introduce the zeroth, first and second order moments of the Wigner function, respectively the number density $n$, the fluid velocity ${\bf u}$ and the pressure dyad ${\bf P}$, defined according to
\begin{eqnarray}
\label{e3}
n &=& \int d^{3}p\,f \,, \\
\label{e4}
m n {\bf u} &=& \int d^{3}p\,{\bf p}\,f \,, \\
\label{e5}
{\bf P} &=& \frac{1}{m}\int d^{3}p\,{\bf p}\otimes{\bf p}\,f - m n {\bf u}\otimes{\bf u} \,.
\end{eqnarray}
It is also convenient to define a third order moment in a component-wise manner, 
\begin{equation}
Q_{ijk} = \frac{1}{m^2}\int d^{3}p\,(p_i - m u_i)(p_j - m u_j)(p_k - m u_k)\,f \,.
\end{equation}

Integrating all terms of Eq. (\ref{e1}) over momenta assuming appropriate boundary conditions yields
\begin{equation}
\label{e6}
\frac{\partial n}{\partial t} + \nabla\cdot(n{\bf u}) = 0 \,,
\end{equation}
which is the continuity equation. Multiplying all terms in Eq. (\ref{e1}) by ${\bf p}/m$, integrating over momenta and using the continuity equation gives
\begin{eqnarray}
\label{e7}
\left(\frac{\partial}{\partial t} + {\bf u}\cdot\nabla\right){\bf u} = \!\!\!\!&\strut&\!\!\!\! - \frac{\nabla\cdot{\bf P}}{m n}    - \frac{e{\bf E}}{m}  \\ &-& \frac{1}{2 m n}\int d^{3}p\,d^{3}r\,d^{3}q\,e^{-i{\bf r}\cdot{\bf q}/\hbar}\,\frac{\partial V({\bf r})}{\partial{\bf r}}\,f\left({\bf x}\!-\!\frac{\bf r}{2}, {\bf p}\!+\!\frac{\bf q}{2}, t\right)\!f\left({\bf x}\!-\!\frac{\bf r}{2}, {\bf p}\!-\!\frac{\bf q}{2}, t\right) \,. \nonumber
\end{eqnarray}
It is important to notice that if the analysis is restricted to a 10-moment model (in terms of $n, {\bf u}$ and the components of the the symmetric dyad $P_{ij}$, then the last term in Eq. (\ref{e1}) with the second-order derivative $\partial^{2}_{\partial p_i\partial p_j}$ would be entirely washed out. Indeed, it does not contribute to Eq. (\ref{e7}) since 
\begin{equation}
\int d^{3}p\, p_k\, \frac{\partial^2 A_{ij}}{\partial p_i \partial p_j} = - \delta_{ik}\int d^{3}p \, \frac{\partial A_{ij}}{\partial p_j} = 0 \,,
\end{equation}
assuming surface terms do not contribute for quantities $A_{ij}$ satisfying decaying boundary conditions in momentum space. Therefore the above mentioned last term in Eq. (\ref{e1}) can contribute only in a higher-order moments hierarchy model.

For completeness, the evolution equation for the second order moment (the energy transport equation) is also presented, as obtained after multiplying all terms in Eq. (\ref{e1}) by $p_i p_j$ and integrating over momentum space,
\begin{eqnarray}
\label{e8}
\Bigl(\frac{\partial}{\partial t} &+& {\bf u}\cdot\nabla\Bigr)P_{ij} = - P_{ki}\frac{\partial u_j}{\partial x_k} - P_{kj}\frac{\partial u_i}{\partial x_k} - P_{ij}\nabla\cdot{\bf u} - \frac{\partial Q_{ijk}}{\partial x_k} \\
&\strut& - \frac{1}{2m}\int d^{3}p\,d^{3}r\,d^{3}q\,e^{-i{\bf r}\cdot{\bf q}/\hbar}\left[\left(p_i - m u_i({\bf x},t)\right) \frac{\partial V({\bf r})}{\partial r_j} + \left(p_j - m u_j({\bf x},t)\right) \frac{\partial V({\bf r})}{\partial r_i}\right] \times \nonumber \\ &\strut& \quad\quad\quad\quad\quad\quad \times\,\, f\left({\bf x}\!-\!\frac{\bf r}{2}, {\bf p}\!+\!\frac{\bf q}{2}, t\right)\!f\left({\bf x}\!-\!\frac{\bf r}{2}, {\bf p}\!-\!\frac{\bf q}{2}, t\right)  - \frac{i\hbar}{8m}\int d^{3}p\,d^{3}r\,d^{3}q\,e^{-i{\bf r}\cdot{\bf q}/\hbar}  \times \nonumber \\ &\strut& \!\!\!\!\!\!\!\!\!\!\!\!\!\!\!\!\!\!\!\times\Bigl\{\frac{\partial V({\bf r})}{\partial r_i}\Bigl[\frac{\partial}{\partial x_j}\!f\!\left({\bf x}\!-\!\frac{\bf r}{2},{\bf p}\!-\!\frac{\bf q}{2},t\right)\!\!f\!\left({\bf x}\!-\!\frac{\bf r}{2},{\bf p}\!+\!\frac{\bf q}{2},t\right)\!\!-\!\!f\!\left({\bf x}\!-\!\frac{\bf r}{2},{\bf p}\!-\!\frac{\bf q}{2},t\right)\!\!\frac{\partial}{\partial x_j}\!f\!\left({\bf x}\!-\!\frac{\bf r}{2},{\bf p}\!+\!\frac{\bf q}{2},t\right)\Bigr] \nonumber \\
 &\strut& \!\!\!\!\!\!\!\!\!\!\!\!\!\!\!\!\!\!\!+\frac{\partial V({\bf r})}{\partial r_j}\Bigl[\frac{\partial}{\partial x_i}\!f\!\left({\bf x}\!-\!\frac{\bf r}{2},{\bf p}\!-\!\frac{\bf q}{2},t\right)\!\!f\!\left({\bf x}\!-\!\frac{\bf r}{2},{\bf p}\!+\!\frac{\bf q}{2},t\right)\!\!-\!\!f\!\left({\bf x}\!-\!\frac{\bf r}{2},{\bf p}\!-\!\frac{\bf q}{2},t\right)\!\frac{\partial}{\partial x_i}\!f\!\left({\bf x}\!-\!\frac{\bf r}{2},{\bf p}\!+\!\frac{\bf q}{2},t\right)\Bigr]\Bigr\} . \nonumber
\end{eqnarray}
If the exchange terms are omitted, then Eqs. (\ref{e6})-(\ref{e8}) reproduce previous results on electrostatic plasma \cite{PLA} but now disregarding quantum diffraction.

\section{Ion-acoustic waves from exchange-fluid theory}

For the fluid treatment of ion-acoustic waves, we have to add the ions continuity equation,
\begin{equation}
\label{ice}
\frac{\partial n_i}{\partial t} + \nabla\cdot(n_i{\bf u}_i) = 0 \,,
\end{equation}
the ions force equation, 
\begin{equation}
\label{ife}
\left(\frac{\partial}{\partial t} + {\bf u}_i\cdot\nabla\right){\bf u}_i =  \frac{e {\bf E}}{m_i} \,,
\end{equation}
and Poisson's equation, 
\begin{equation}
\label{pe}
\nabla\cdot{\bf E} = \frac{e}{\varepsilon_0}(n_i - n) \,.
\end{equation}
We have denoted $n_i$ as the ions number density, ${\bf u}_i$ the ions velocity field and $m_i$ the ions mass. Due to their large mass, exchange effects were disregarded for ions, considered to be simply ionized and cold as well. Obviously the ionic fluid equations can be also derived taking the moments of the respective Vlasov equation and assuming a cold velocity shifted Maxwellian distribution function, which is appropriate since ions can be usually taken as classical in view of their larger mass. 

For our purposes, while writing it for completeness, Eq. (\ref{e8}) for the time-evolution of the third-order electrons moment can be ignored. Therefore we have the basic model in terms of Eqs. (\ref{e6}) and (\ref{ice}), resp. the electrons and ions continuity equations, Eqs. (\ref{e7}) and (\ref{ife}), resp. the electrons and ions force equations, and Poisson's equation (\ref{pe}). These are 9 equations for the 9 relevant quantities, namely $n, n_i$, the components of ${\bf u}, {\bf u}_i$, and the scalar potential $\phi$ such that ${\bf E} = - \nabla\phi$. As expected, we have a closure problem since Eq. (\ref{e7}) not only contains the pressure dyad but also the electrons distribution function, which satisfies the exchange kinetic equation. As usual the pressure term can be managed assuming an equation of state compatible with a local thermodynamic equilibrium. We postpone the problem of expressing the exchange term in Eq. (\ref{e7}) in terms of more familiar hydrodynamic variables to future works.  Nevertheless, we can provide a recipe to deal with the exchange term in the case of linear waves, as follows. 

Exchange effects on linear ion-acoustic waves can be investigated assuming 
\begin{eqnarray}
n &=& n_0 + \delta n \,, \quad n_i = n_0 + \delta n_i \,, \quad \phi = \delta\phi \,, \nonumber \\
{\bf u} &=& \delta{\bf u} \,, \quad {\bf u}_i = \delta{\bf u}_i \,, \quad f = f_0({\bf p}) + \delta f({\bf p}) \,,
\end{eqnarray}
where the $\delta$ identify first order quantities all of them proportional to a plane wave $\exp[i({\bf k}\cdot{\bf x} - \omega t)]$. 

Retaining only first order terms we easily find 
\begin{equation}
\label{w1}
\omega\delta n = n_0 {\bf k}\cdot\delta{\bf u} \,, \quad \omega\delta n_i = n_0 {\bf k}\cdot\delta{\bf u}_i \,, \quad \omega\delta{\bf u}_i = \frac{e{\bf k}\delta\phi}{m_i} \,, \quad k^2\delta\phi = \frac{e}{\varepsilon_0}(\delta n_i - \delta n) 
\end{equation}
where $k = |{\bf k}|$ and 
\begin{eqnarray}
0 &=& - \frac{m_i c_s^2 {\bf k} \delta n}{n_0} +   e{\bf k}\delta\phi \label{x} \\
&-& \frac{\hbar e^2}{2n_0\varepsilon_0} \int d^3 p\, d^3 q \frac{({\bf q} + \hbar{\bf k}/2)}{|{\bf q} + \hbar{\bf k}/2|^2}\left[f_{0}\left({\bf p}+\frac{\bf q}{2}\right)\,\delta f\left({\bf p}-\frac{\bf q}{2}\right) + f_{0}\left({\bf p}-\frac{\bf q}{2}\right)\,\delta f\left({\bf p}+\frac{\bf q}{2}\right)\right] \,, \nonumber
\end{eqnarray}
where the inertialess electrons approximation is used in view of the low frequency assumption and $P_{ij} = p(n)\delta_{ij}$ which is valid for isotropic equilibria. The ion-acoustic speed $c_s$ comes from
\begin{equation}
c_s^2 = \frac{1}{m_i}\left(\frac{dp}{dn}\right)_{n=n_0} \,.
\end{equation}
In view of the neglect of quantum diffraction (discussed in more detail in Section IV) it is possible to expand the integrand in Eq. (\ref{x}) for small $k$ to obtain 
\begin{equation}
0 = - \frac{m_i c_s^2 {\bf k} \delta n}{n_0} +   e{\bf k}\delta\phi \label{yyy} - \frac{\hbar^2 e^2}{n_0\varepsilon_0} \int d^3 p\, d^3 q 
\left(\frac{\bf k}{2q^2} - \frac{{\bf k}\cdot{\bf q}\,{\bf q}}{q^4}\right)
f_{0}\left({\bf p}+\frac{\bf q}{2}\right)\,\delta f\left({\bf p}-\frac{\bf q}{2}\right)  \,.
\end{equation}
The supposed leading order term (with ${\bf k}=0$) does not contribute in view of parity properties. 

As mentioned earlier, there is an extra closure problem since the exchange term in the electrons momentum equation involves the electrons probability distribution function, whose variation $\delta f$ therefore appears in Eq. (\ref{yyy}). To deal with the question, notice that the exchange term is a correction term, an hypothesis to be confirmed at the end of the calculation. Hence we can use just the linearized Vlasov  equation without exchange term to derive in the static approximation 
\begin{equation}
\frac{{\bf k}\cdot{\bf p}}{m}\,\delta f({\bf p}) + e{\bf k}\cdot\frac{\partial f_0({\bf p})}{\partial{\bf p}}\,\delta\phi \simeq 0 \,,
\end{equation}
which can be further simplified in the case of isotropic equilibria, 
\begin{equation}
\label{df}
f_0 = f_{0}(\epsilon) \,, \quad \epsilon = \frac{p^2}{2m} \quad \Rightarrow \quad \delta f({\bf p}) = - e\,\frac{df_0}{d\epsilon}\delta\phi \,.
\end{equation}

Inserting Eq. (\ref{df}) into Eq. (\ref{yyy}) the result is 
\begin{equation}
1 + \chi_e(\omega,{\bf k}) + \chi_i(\omega,{\bf k})  = 0 \,,
\end{equation}
where 
\begin{equation}
\label{xx}
\chi_e(\omega,{\bf k}) = \frac{e}{\varepsilon_0 k^2}\frac{\delta n}{\delta\phi} = \frac{\omega_{pi}^2}{c_s^2k^2}\left[1 + \frac{\hbar^2 e^2}{n_0\varepsilon_0} \int\! d^3 p\, d^3 q 
\left(\frac{1}{2q^2} - \frac{({\bf k}\cdot{\bf q})^2}{k^2 q^4}\right)
f_0(\epsilon_{+})\left(\frac{df_0}{d\epsilon}\right)_{\!\!\epsilon_{-}}\right] 
\end{equation}
is the electrons susceptibility with $\omega_{pi}^2 = n_0 e^2/m_i\varepsilon_0$ and 
\begin{equation}
\epsilon_{\pm} = \frac{({\bf p} \pm {\bf q}/2)^2}{2m} 
\end{equation}
and  where  
\begin{equation}
\chi_i(\omega,{\bf k}) = - \frac{e}{\varepsilon_0 k^2}\frac{\delta n_i}{\delta\phi} =   - \frac{\omega_{pi}^2}{\omega^2}
\end{equation}
is the ions susceptibility. 

The result up to now is valid for isotropic electronic equilibrium distribution functions in general, within the assumed validity conditions.
The most prominent case for the evaluation of exchange effects would be the Fermi-Dirac equilibrium, but for analytical reasons here we follow the trend of \cite{Zamanian} and consider a Maxwell-Boltzmann equilibrium, 
\begin{equation}
f_0(\epsilon) = \frac{n_0}{(2\pi m \kappa_B T)^{3/2}}\,\exp\left(-\,\frac{\epsilon}{\kappa_B T}\right) \,.
\end{equation}

After a simple algebra one has 
\begin{equation}
\label{chif}
\chi_e = 
\frac{\omega_{pi}^2}{c_s^2 k^2}\left[1 - \frac{1}{12}\left(\frac{\hbar\omega_{p}}{\kappa_B T}\right)^2\right] \,,
\end{equation}
where $\omega_{p}^2 = n_0 e^2/m\varepsilon_0$, and $p = n \kappa_B T$ so that $c_s^2 = \kappa_B T/m$.

To first order in the exchange effects, one has 
\begin{equation}
\omega^2 = \frac{\omega_{pi}^2c_s^2 k^2}{\omega_{pi}^2 + c_s^2 k^2}\left[1 + \frac{1}{12}\left(\frac{\hbar\omega_{p}}{\kappa_B T}\right)^2
\,\,\frac{\omega_{pi}^2}{\omega_{pi}^2 + c_s^2 k^2}\right] \,.
\end{equation}

Ion-acoustic waves in the quasi-neutral regime $c_s k \ll \omega_{pi}$ reduce to 
\begin{equation}
\label{eiaw}
\omega = c_s k\left[1 + \frac{1}{24}\left(\frac{\hbar\omega_{p}}{\kappa_B T}\right)^2\right] \,,
\end{equation}
while ionic waves ($c_s k \gg \omega_{pi}, \omega = \omega_{pi}$) have exchange effects only at a higher order.

\section{Validity conditions}

We have made a number of assumptions to be detailed, as follows.

\begin{figure}
\begin{center}
\includegraphics[width=4.5in]{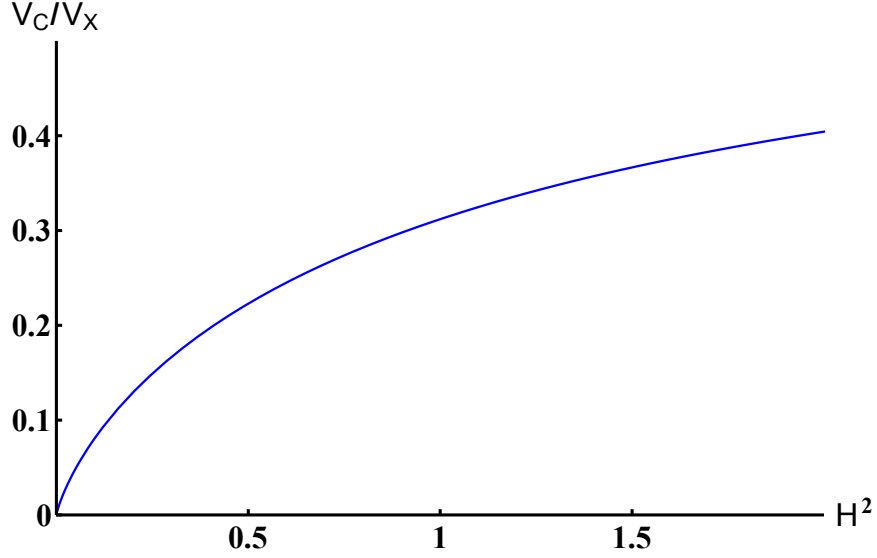}
\caption{Relative strength of correlation effects compared to exchange effects using the empirical DFT functionals, from Eq. (\ref{vc}). }
\label{fig1} 
\end{center}
\end{figure}

\begin{enumerate}
\item Neglect of quantum diffraction. Besides exchange effects whose origin is the fermionic statistics alone, quantum undulatory effects (quantum diffraction) can also be a relevant quantum effect. These last have been disregarded from the beginning, since the derivation of the exchange kinetic theory have taken into account only the anti-symmetry of the two-body electron particle distribution function in the otherwise classical 
BBGKY (Bogoliubov–Born–Green–Kirkwood–Yvon) hierarchy. To estimate the quantum diffraction and disregarding irrelevant numerical factors, we consider the dispersion relation of quantum ion-acoustic waves without exchange correction \cite{MF, H}, where quantum diffraction comes from the Bohm potential term, providing the condition 
\begin{equation}
\label{nqd}
k^2 c_s^2 \gg \frac{\hbar^2 k^4}{m_e m_i} \quad \Rightarrow \quad \hbar k \ll m v_T \,,
\end{equation}
which is the long wave-length approximation implicit in the step from Eq. (\ref{x}) to Eq. (\ref{yyy}), where $v_T = \sqrt{\kappa_B T/m}$.

\item Non-degenerate assumption: $\kappa_B T >> \epsilon_F$, where $\epsilon_F = (\hbar^2/2m)(3\pi^2 n_0)^{2/3}$ is the Fermi energy. This assumption holds only for the choice of a Maxwellian equilibrium for the sake of simplicity.

\item Ideality. Since we have started from the exchange kinetic theory which at the present does not take into account collisions, we must have a small graininess parameter $g$, which is the ratio between average electrostatic and kinetic energies. For a Maxwellian plasma, we have 
\begin{equation}
g = \frac{e^2}{4\pi\varepsilon_0 r_S \kappa_B T} \ll 1 \,, \quad \frac{4\pi r_S^3}{3} = \frac{1}{n_0} \,,
\end{equation}
where $r_S$ is the Wigner-Seitz radius. We find
\begin{equation}
\label{g}
\left(\frac{\hbar\omega_p}{\kappa_B T}\right)^2 = 4 \left(\frac{2}{3\pi^2}\right)^{1/3} \frac{\epsilon_F}{\kappa_B T}\,\,g \,,
\end{equation}
which explains why the exchange effects should be just a correction, at least for non-strongly coupled and non-degenerate plasmas. 

\item Neglect of Landau damping. Beyond dogmas, the concrete reasons for the need of kinetic theory arise when the simpler, fluid model are not capable of describing some relevant aspect addressed in kinetic theory. One such effect is Landau damping. Ion-acoustic waves in classical plasma described by the Vlasov-Poisson system have a negligible collisionless damping provided $m \ll m_i$ and $T_i \ll T$, where $T_i$ is the ions temperature \cite{Boyd}. Ref. \cite{Zamanian} has obtained the Landau damping rate from exchange kinetic theory with an underlying Maxwellian equilibrium for electrons and cold ions and found it is small provided $m/m_i \ll 1$, inline with the inertialess electrons assumption. There the exchange effects provide a small correction to the already small classical Landau damping (as shown in Eq. (16) of \cite{Zamanian}). 

\item Inertialess electrons. The inertialess electrons condition is equivalent to a low frequency assumption \cite{Boyd} given by 
\begin{equation}
\omega \simeq k c_s \ll k v_T \quad \Rightarrow \quad m \ll m_i \,.
\end{equation}
This condition is attained for heavy ions, the same rule for disregarding Landau damping of the ion-acoustic wave. For hydrogen plasma one has $c_s/v_F = 0.02$.

\item Neglect correlations. It has become popular \cite{Crouseilles}, \cite{Xia}-\cite{Zeba} to investigate exchange-correlation effects in plasmas using effective empirical potentials directly taken from equilibrium density functional theory (DFT). At the moment a quantum kinetic theory for the full exchange-correlation effects is not available. Hence a fluid theory for exchange-correlation from first principles is also not available, only exchange effects have been presently taken into account. Nevertheless for the sake of an estimate it is possible to measure the relevance of correlation effects using the DFT functionals $V_C$ (for correlations) and $V_X$ (for exchange), using Eqs. (10) and (11) of \cite{Crouseilles} at equilibrium ($n = n_0$), yielding 
\begin{equation}
\label{vc}
\frac{V_C}{V_X} = 0.25\,H^2\,\ln\left(1 + \frac{2.52}{H^2}\right) < 0.63 \,,
\end{equation}
where $H = \hbar\omega_p/m v_F^2$ and $v_F = (2\epsilon_F/m)^{1/2}$ is the Fermi velocity. When $H \gg 1$ one has $V_C/V_X \simeq 0.63$, within the accuracy of the empirical DFT functionals, but in the context of an ideal degenerate plasma one always has $H \ll 1$. Figure (\ref{fig1}) shows the ratio between $V_C$ and $V_X$ which is a function of $H^2$ only. The estimate shows that exchange effects tend to be dominant, although not directly related to a Maxwellian plasma as discussed here. Nevertheless in a rough translation the parameter $H^2$ in a degenerate plasma corresponds to the graininess parameter $g$ in a non-degenerate plasma, since it is the ratio between the average electrostatic and kinetic energies in fully degenerate plasma. Hence Eq. (\ref{vc}) can be read as a function of  $g$. In this picture, $V_C/V_X \rightarrow 0$ the more ideal the plasma is, which is not surprising.  

\end{enumerate}

\section{Comparison with exchange kinetic theory}

The linear dispersion relation for ion-acoustic waves described by the exchange kinetic theory was previously obtained in \cite{Zamanian}. However, it is relevant to analyze the kinetic approach again, to have a detailed comparison with the fluid approach. Consider the electrons susceptibility
\begin{equation}
\chi_e = \frac{e}{\varepsilon_0 k^2}\int d^3 p \,\frac{\delta f}{\delta\phi} \,.
\end{equation}
Using Eq. (\ref{e1}) in the static limit ($\omega \ll k v_T$), for isotropic equilibrium, with a small $k$ expansion like for Eq. (\ref{yyy}), 
and approximating $\delta f$ in the exchange term by the Vlasov expression as in Eq. (\ref{df}), the result is  
\begin{equation}
\chi_e = \frac{\omega_{pi}^2}{c_s^2 k^2}\left[1 - \frac{\hbar^2 m e^4 c_s^2}{\varepsilon_0^2 \omega_{pi}^2}\int \frac{d^3 p}{{\bf k}\cdot{\bf p}}\,\frac{\partial}{\partial{\bf p}}\bullet \int d^3 q \left(\frac{\bf k}{2q^2} - \frac{{\bf k}\cdot{\bf q}\,{\bf q}}{q^4}\right) f_0(\epsilon_+)\left(\frac{df_0}{d\epsilon}\right)_{\epsilon_{-}}\right] \,. 
\label{ab2}
\end{equation}
With the Maxwellian equilibrium this can be analytically calculated yielding 
\begin{equation}
\label{chik}
\chi_e = 
\frac{\omega_{pi}^2}{c_s^2 k^2}\left[1 - \frac{1}{6}\left(\frac{\hbar\omega_{p}}{\kappa_B T}\right)^2\right] \,,
\end{equation}
which is the same as the exchange fluid result (\ref{chif}) except for a factor $2$ in front of the exchange contribution. Proceeding with the same ions susceptibility as before and in the quasi-neutral approximation one find 
\begin{equation}
\label{eeiaw}
\omega = c_s k\left[1 + \frac{1}{12}\left(\frac{\hbar\omega_{p}}{\kappa_B T}\right)^2\right] \,,
\end{equation}
to be compared with Eq. (\ref{eiaw}) for the exchange fluid dispersion relation. 

In Ref. \cite{Zamanian}, the exchange correction for ion acoustic waves in a Maxwellian background was derived from kinetic theory without applying the static limit ($\omega \ll k v_T$). For this reason, the dispersion relation was obtained only after a numerical integration, while here Eq. (\ref{ab2}) can be analytically solved. Equation (16) in \cite{Zamanian} shows $\omega = c_s k [1 + 0.8 (\hbar\omega_p/\kappa_B T)^2]$, omitting the small Landau damping term, while Eq. (\ref{eeiaw}) shows a numerical factor $1/12 = 0.08$ for the exchange correction. 

For the sake of comparison, we see that the exchange fluid electrons susceptibility from Eq. (\ref{xx}) has a different expression for the exchange correction, as for the exchange kinetic general result in Eq. (\ref{ab2}). This is the origin of the factor $2$ difference appearing in Eqs. (\ref{eiaw}) and (\ref{eeiaw}) resp. for the fluid and kinetic exchange contributions. One more source of discrepancy is the $\partial^2/\partial p_i \partial p_j$ term in the exchange contribution in the kinetic equation (\ref{e1}), which is immediately washed out in a low order moments approach. In the static limit, this term can be shown to have no impact in the exchange kinetic electrons susceptibility. However, it can be significant for fast waves other than slow waves such as the ion-acoustic branch. 

\section{Conclusion}

In this work, we have developed the macroscopic, fluid theory arising after taking moments from the recently introduced first principles exchange kinetic theory \cite{Zamanian}. The model is then applied to the case of linear ion-acoustic waves. Closure is obtained assuming that the exchange effects are a correction, which is verified at the end of the calculation in view of the underlying collisionless and long wavelength approximations. The analytical expression have been obtained for general isotropic equilibria and developed in all detail in the case of a Maxwellian equilibrium. The results were compared with those from exchange kinetic theory and their validity conditions discussed. At the same time, these validity conditions show natural issues in order to generalize the available kinetic and fluid exchange models. Alternative closure methods will be reported soon. The exchange-fluid equations can be an useful tool to investigate nonlinear aspects of quantum plasmas where exchange effects are prominent. 

\newpage
\textbf{Acknowledgements} \par 
The author acknowledges financial support by CNPq (Conselho Nacional de De\-sen\-vol\-vi\-men\-to Cient\'{\i}fico e Tecnol\'ogico), Brazil.


\begin{thebibliography}{99}
\bibitem{Bloch} F. Bloch, Zeits.f. Phys. {\bf 57}, 545 (1929). 
\bibitem{Eliezer} S. Eliezer, A. Ghatak and H. Hora, {\it Fundamentals of Equations of State} (Singapore, World Scientific, 2002). 
\bibitem{Zamanian} J. Zamanian, M. Marklund and G. Brodin, Phys. Rev. E {\bf 88}, 063105 (2013).
\bibitem{Ekman} R. Ekman, J. Zamanian and G. Brodin, Phys. Rev. E {\bf 92}, 013104 (2015).
\bibitem{Roos} O. von Roos and J. S. Zmuidzinas, Phys. Rev. {\bf 121}, 941
(1961).
\bibitem{Kana}  H. Kanazawa, S. Misawa, and K. Fujita, Progr. Theoret. Phys.
{\bf 23}, 426 (1960).
\bibitem{Nozieres} P. Nozieres and D. Pines, Phys. Rev. {\bf 111}, 442 (1958).
\bibitem{Zamanian2} J. Zamanian, M. Marklund and G. Brodin, Eur. Phys. J. D {\bf 69}, 25 (2015).
\bibitem{Grad} H. Grad, Commun. Pure Appl. Math. {\bf 2}, 331 (1949); E. Siregar and M. L.
Goldstein, Phys. Plasmas {\bf 3}, 1437 (1996); P. Goswami, T. Passot
and P. L. Sulem, Phys. Plasmas {\bf 12}, 102109 (2005).
\bibitem{PLA} F. Haas, M. Marklund, G. Brodin and J. Zamanian, Phys. Lett. A. {\bf 374}, 481 (2010). 
\bibitem{NJP} F. Haas, J. Zamanian, M. Marklund and G. Brodin, New J. Phys. {\bf 12}, 073027 (2010).
\bibitem{Crouseilles} N. Crouseilles, P.-A. Hervieux and G. Manfredi, Phys. Rev. B {\bf 78}, 155412 (2008).
\bibitem{Brey} L. Brey, J. Dempsey, N. F. Johnson and B. I. Halperin, Phys. Rev. B {\bf 42}, 124 (1990).
\bibitem{Xia} H. Cai-Xia and X. Ju-Kui, Chin. Phys. B {\bf 22}, 025202 (2013).
\bibitem{Ourabah} K. Ourabah and M. Tribeche, Phys. Rev. E {\bf 88}, 045101 (2013). 
\bibitem{Akbari} M. Akbari-Moghanjoughi and P. K. Shukla, Phys. Rev. E {\bf 86}, 066401 (2012). 
\bibitem{Zeba} I. Zeba, M. E. Yahia, P. K. Shukla and W. M. Moslem, Phys. Lett. A {\bf 376}, 2309 (2012). 
\bibitem{Brodin2019} G. Brodin, R. Eckman and J. Zamanin, Phys. Plasmas {\bf 26}, 092113 (2019).
\bibitem{MF} G. Manfredi and F. Haas, Phys. Rev. B {\bf 64}, 075316 (2001).
\bibitem{H} F. Haas, L. G. Garcia, J. Goedert and G. Manfredi, Phys. Plasmas {\bf 10}, 3858 (2003). 
\bibitem{Boyd} T. J. M. Boyd and J. J. Sanderson, {\it The Physics of Plasmas} (Cambridge, New York, 2003). 
\end{thebibliography}
\end{document}